

\def\header#1{{\bf #1} \nobreak}

\def\proclaim#1{{\vskip 2mm \goodbreak \hbox{\bf #1} \vskip 1mm}}
\def\endproclaim{\vskip 2 mm}
\def\demo#1{{\it#1: }}
\def\enddemo{\vskip 2 mm}
\def\qed{\hfill{\vrule height 3mm width 2mm depth 0mm}}

\def\ref#1{{${}^{#1}$}}

\def\:{\hbox{\kern-0.03em:\kern-0.03em}}
\def\ddot{{\kern-0.04em\cdot\kern-0.04em}}
\def\mxn{{m \leftrightarrow n}}

\def\Reals{{\bf R}}
\def\Ints{{\bf Z}}
\def\oj{{\bf g}}

\def\e{\hbox{e}}
\def\emx{\e^{m \ddot x}}
\def\enx{\e^{n \ddot x}}
\def\emnx{\e^{(m+n) \ddot x}}
\def\enmx{\e^{-m \ddot x}}

\def\args#1#2{{{#1_1} \ldots {#1_{#2}}}}
\def\argsx#1#2#3{{{#1_1} \ldots #3 \ldots {#1_{#2}}}}

\def\sigmas{\args \sigma p}
\def\taus{\args \tau q}

\def\sigmasmu{\argsx \sigma p \mu}

\def\tausnu{\argsx \tau q \nu}

\def\vacuum{{| \kern-0.04em 0 \kern-0.04em \rangle}}

\def\notrace{(p-q)/(N+1)}

\def\CC{{\bf C}}
\def\TT{{\bf T}}
\def\ii{{\bf k}}
\def\ll{{\bf l}}

\def\d{\partial}

\def\endchapter{{\vglue 15 mm} \goodbreak}
\def\newpage{{\vfill \eject}}
\baselineskip = 12pt
\overfullrule = 0pt
\magnification = 1200

\centerline{\bf Conformal fields: a class of representations of Vect(N)}

\vglue 30pt

T A Larsson
\footnote*{Supported by the Swedish Natural Science Research Council (NFR)}

Department of Theoretical Physics,

Royal Institute of Technology

100 44 Stockholm, Sweden

email: tl@theophys.kth.se

\vglue 30pt
(October 1991)
\vglue 60pt

\header{Abstract}

$Vect(N)$, the algebra of vector fields in $N$ dimensions, is studied.
Some aspects of local differential geometry are formulated as $Vect(N)$
representation theory.
There is a new class of modules, {\it conformal fields},
whose restrictions to the subalgebra
$sl(N+1) \subset Vect(N)$ are finite-dimensional $sl(N+1)$
representations. In this regard they are simpler than tensor fields.
Fock modules are also constructed. Infinities, which are unremovable even by
normal ordering, arise unless bosonic and fermionic degrees of freedom match.

\vglue 30 pt
PACS number: 11.30

\newpage

\header{1. Introduction}

It seems to be a reasonable assumption that the physical content of a
theory is independent of the choice of coordinate system. This leads by
definition to
the conclusion that any physical quantity must be an intrinsic object in
the sense of differential geometry, i.e. that it must transform as a
representation of the group $Diff(M)$ of all coordinate transformations,
or diffeomorphisms, on the base manifold $M$. Something which is not a
representation of $Diff(M)$ is clearly an artefact of the choice of
coordinate system and thus unsuitable for physics.

Diffeomorphisms on arbitrary manifolds have been studied by mathematicians
from various points of view for a long time\ref{1-6}, and they have
recently attracted some interest by physicists as natural generalizations of
the one-dimensional case\ref{7-12}. Nevertheless,
the representation theory of $Diff(M)$
is not very well understood when $\dim M > 1$. In particular, no
irreducible or lowest-weight representations are known to us\ref{13}.
In order to study the diffeomorphism group it is reasonable to start with its
Lie algebra of vector fields on $M$. Moreover, we
only consider vector fields locally; we expect problems of
homological nature to appear in a global approach, but an understanding of
the local properties is certainly a prerequisite for global analysis.
Such a program has of course been carried out in great detail on the circle,
where it leads to the Virasoro algebra, i.e. the universal central extension
of $Vect(1)$\ref{13-15}. The same algebra is also at the core of conformal
field theory which is important to the theory of critical phenomena in two
dimensions\ref{16}.

In section 2 we define $Vect(N)$, the algebra of vector fields in
$N$-dimensional space.
The algebra is given in a plane wave basis; this is always possible to
do locally.
We also show that it does not admit any central extension except when $N = 1$.
Section 3 is the main part of this paper. We begin by formulating local
differential geometry (tensor fields, forms and exterior derivatives) as
$Vect(N)$ representation theory.
Tensor fields are intimately related to tensors of $gl(N) \subset Vect(N)$.
However, the largest finite-dimensional subalgebra is not $gl(N)$ but
$cl(N) \cong sl(N+1) \subset Vect(N)$, which is obtained from $gl(N)$ by
adding translations and conformal transformations. Any $Vect(N)$ module
yields of course an $sl(N+1)$ module by restriction, but tensor fields do
not correspond to finite-dimensional $sl(N+1)$ representations. We construct
{\it conformal fields}, which are $Vect(N)$ modules having this desirable
property, and initiate their study.
Since conformal fields are more natural than tensor fields, at least regarding
their conformal properties, they could be a good tool for interesting physics.
Section 4 contains a brief discussion of Fock modules. It is found that
infinities can not be removed by normal ordering, but they can be cancelled
between bosons and fermions.
In the final section it is noted that any representation of $Vect(N)$ gives
rise to a representation of the Poisson algebra in $N$ dimensions. Using
the results of section 3, a host of Poisson modules can thus be written down.

\endchapter

\header{2. Definition of $Vect(N)$}

Select a point on a manifold $M$ such that $\dim M = N$. In some
neighborhood of this point we introduce local coordinates
$x = (x_1, \ldots, x_N)$, where the origin is the selected point.
For simplicity we only consider
coordinate patches that are hypercubic in the sense that any function can be
expanded in a plane wave basis. A basis for the function algebra is thus
given by $\{\e^{im \ddot x}\}_{m \in \Lambda}$,
where $m = (m^1, \ldots, m^N) \in \Lambda \subset \Reals^N$ is a point
of an $N$-dimensional lattice and
$m \ddot x \equiv m^\mu x_\mu$. The summation convention is implied unless
otherwise stated.

A vector field locally has the form $f(x) \d^\mu$, which in this region is
a linear combination of the basis elements
$$ L^\mu(m) = -i \exp(im \ddot x) \d^\mu.
	\eqno(2.1) $$
It is immediate to check that these generators obey the Lie algebra
$Vect(N)$, with basis $\{L^\mu(m)\}_{m \in \Lambda}$ and brackets
$$ [L^\mu(m), L^\nu(n)] = n^\mu L^\nu(m+n) - m^\nu L^\mu(m+n),
	\eqno(2.2) $$
where $\mu, \nu = 1, \ldots, N = \dim \Lambda$, and $m, n \in \Lambda$.
This algebra is sometimes referred to as a generalized Witt algebra [6].
By means of the dual lattice,
$$\Lambda^* = \{\alpha = (\alpha_1, \ldots, \alpha_N):
m \ddot \alpha \in 2\pi \Ints\},
	\eqno(2.3) $$
the neighboorhood can be described
as the torus $\Reals^N / \Lambda^*$. All considerations in this paper
are local, but on this torus the results hold globally.

By a complex rescaling of the lattice $\Lambda$, (2.1) can be brought
to the form
$$ L^\mu(m) = \emx \d^\mu,
	\eqno(2.4) $$
which also satisfies (2.2). Since only algebraic properties are of
interest in this
paper we take this as the defining representation of $Vect(N)$ and model
other
representations on it. This formalism saves many factors of $i$, but
more conventional expressions can be recovered by restricting
$\Lambda$ to a purely imaginary lattice.

$Vect(N)$ has also another natural realization, as the algebra
of holomorphic vector fields in $N$ dimensions.
$$ L^\mu(m) = (x_1)^{m^1} (x_2)^{m^2} \ldots (x_N)^{m^N} x_\mu
\partial^\mu \qquad \hbox{(no sum on $\mu$).}
	\eqno(2.5) $$
In this case the basis should be restricted to
$\{L^\mu(m) | m^\mu \ge -1$ and
$ m^\nu \ge 0 \; \forall \; \nu \neq \mu \}$,
which ensures that the vector fields
do not have any singularities at the origin. Thus, this is a
basis for vector fields which can be expanded in a Taylor series around the
origin. A point worth noting is that the $N=1$ version of this ``amputated''
algebra is a Virasoro subalgebra for any value of the central
charge, because the central extension only enters in the brackets
$[L(m), L(-m)]$ with $m \ge 2$. Hence any Virasoro module, for arbitrary
$c$, yields a representation of amputated $Vect(1)$ by restriction. This
is of course true only on the level of linear representations; once unitarity
is considered, involution brings back the central charge.

We will henceforth focus on representations modelled on (2.4),
but most results can be translated to the second kind (2.5)
by means of the following dictionary.
$$\emx \to \prod_\mu (x_\mu)^{m^\mu},
\qquad \qquad
\d^\mu \to x_\mu \d^\mu \qquad \hbox{(no sum on $\mu$)}, $$
$$m \ddot x \to \sum_\mu m^\mu \log(x_\mu),
\qquad \qquad
x_\mu \to \log(x_\mu).  \qquad {}
	\eqno(2.6) $$

At this point, there is an obvious question for anyone acquinted to the
Virasoro algebra. The answer to this question is negative;
$Vect(N)$ does not admit any non-trivial central extension except when
$N = 1$.
A proof of this was given already in Ref. 9, but we give the argument
here for convenience. Look for a central extension of the form
$f^{\mu \nu}(m) \delta(m+n)$,
where $f^{\nu \mu}(-m)$ = $-f^{\mu \nu}(m)$ and
$\delta(m)$ is the multi-dimensional Kronecker symbol.
Let $L_k = \alpha \ddot L(km) / \alpha \ddot m$,
$k, l \in \Ints$, $m \in \Lambda$, and $\alpha_\mu$ is a fixed vector.
Then
$$[L_k, L_l] = (l - k) L_{k+l} + {{\alpha \ddot
\alpha \ddot f(km)} \over {(\alpha \ddot m)^2}} \delta((k+l)m),
	\eqno(2.7) $$
so $L_k$ satisfies a central extension of the
Witt algebra. As is well known, any non-trivial such extension must be
cubic. Since this relation
holds for arbitrary choices of $m$ and $\alpha_\mu$, $f^{\mu \nu}(m)$
must in fact
be cubic itself. Make the most general ansatz possible,
$f^{\mu \nu}(m) = c^{\mu \nu}_{\rho \sigma \tau} m^\rho m^\sigma m^\tau
\equiv m \ddot m \ddot m \ddot c^{\mu \nu}$, where the coefficient is
separately
symmetric in its three lower and two upper indices. The Jacobi identities
yield the conditions
$$ n^\nu m \ddot m \ddot m \ddot c^{\mu \sigma} + n^\sigma m \ddot m
\ddot m \ddot c^{\mu \nu}
+ n^\mu m \ddot m \ddot m \ddot c^{\sigma \nu} = 3 m^\nu m \ddot m \ddot
n \ddot c^{\sigma \mu}
	\eqno(2.8) $$
$$ m^\nu m \ddot n \ddot n \ddot c^{\sigma \mu} = n^\mu m \ddot m \ddot
n \ddot c^{\sigma \nu},
	\eqno(2.9) $$
for all $1 \le \mu, \nu, \rho, \sigma, \tau \le N$. These equations hold
identically when $N = 1$,
but do not have a solution otherwise. E.g., in the second equation the LHS is
symmetric in $\mu$ and $\sigma$, whereas the RHS is symmetric in
$\nu$ and $\sigma$. Hence both sides must be symmetric in
all three indices, which clearly is impossible.

We have earlier looked for non-central extensions of $Vect(N)$ which reduce
to the usual Virasoro term when $\dim M = 1$\ref{10, 11},
but we now believe this to be an
uninteresting problem. The reason for this is that it seems to be no natural
way to generate non-central extensions in Fock modules, and therefore we
doubt that the non-central terms have anything to do with representation
theory. This
point will be elaborated in section 4, where we show that normal ordering
gives rise to infinite central extensions rather than to non-central ones.

Ragoucy and Sorba\ref{12} have recently discussed central extensions of
current algebras in $N$ dimensions, i.e. higher-dimensional Kac-Moody
algebras.
$Vect(N)$ arises in this context as derivations of these algebras.
However, not all vector fields are compatible with their extensions,
which was stressed in Ref. 11.
Our point of view is therefore complimentary to theirs: they look for central
extensions and reject vector fields which are not compatible with these,
whereas
we consider arbitrary vector fields and hence reject central extensions.

\endchapter

\header{3. Conformal fields}

It is straight-forward to formulate most aspects of differential geometry
as representation theory of $Vect(N)$. For example, a tensor field is a
$Vect(N)$ module, constructed as follows.
If $T^\mu_\nu$ is a $gl(N)$ generator, i.e.
$$ [T^\mu_\nu, T^\sigma_\tau] = \delta^\mu_\tau T^\sigma_\nu
- \delta^\sigma_\nu T^\mu_\tau,
	\eqno(3.1) $$
then
$$ L^\mu(m) = \emx (\d^\mu + m \ddot T^\mu)
	\eqno(3.2) $$
satisfies $Vect(N)$. This is proved by direct computation:
$$ \eqalign{
[L^\mu(m), L^\nu(n)] &= [\emx (\d^\mu + m \ddot T^\mu),
\enx (\d^\nu + n \ddot T^\nu)] \cr
&=\emnx \big(n^\mu (\d^\nu + n \ddot T^\nu)
 + n^\mu m \ddot T^\mu - \mxn \big) \cr
&= n^\mu \emnx (\d^\nu + (m+n) \ddot T^\nu) - \mxn \cr
&= n^\mu L^\nu(m+n) - \mxn, }
	\eqno(3.3) $$
where $\mxn$ stands for the analogous terms with $m$ and $n$
(and $\mu$ and $\nu$) interchanged.

This observation provides us with a host of $Vect(N)$ representations,
one for
each finite-dimensional $gl(N)$ representation. As is well known
(and easy to verify), there are $gl(N)$ modules $\TT^p_q(\lambda)$
with bases $\upsilon^\sigmas_\taus$ and action
$$ T^\mu_\nu \upsilon^\sigmas_\taus
= \lambda \delta^\mu_\nu \upsilon^\sigmas_\taus
- \sum_{i=1}^p \delta^{\sigma_i}_\nu \upsilon^\sigmasmu_\taus
+ \sum_{j=1}^q \delta^\mu_{\tau_j} \upsilon^\sigmas_\tausnu.
	\eqno(3.4) $$
The action of (3.2) on
$\phi^\sigmas_\taus(x) = \upsilon^\sigmas_\taus \otimes f(x)$
yields the corresponding $Vect(N)$ modules.
$$ \eqalign {
L^\mu(m) &\phi^\sigmas_\taus(x)
= \emx \bigg( (\lambda m^\mu + \d^\mu) \phi^\sigmas_\taus(x) \cr
&\qquad -\sum_{i=1}^p m^{\sigma_i} \phi^\sigmasmu_\taus(x) +
\sum_{j=1}^q \delta^\mu_{\tau_j} m^\nu  \phi^\sigmas_\tausnu(x) \bigg). }
	\eqno(3.5) $$
These modules, which we also denote by $\TT^p_q(\lambda)$,
are called {\it tensor fields}.
There are submodules consisting of symmetric,
skew-symmetric and traceless tensors, etc.

Moreover, the only class of
module homomorphisms connects anti-symmetric tensor fields (forms); this
is the
exterior derivative. The simplest non-tensorial field is the {\it connection},
whose transformation law reads
$$ \eqalign{
L^\mu(m) \Gamma^{\rho \sigma}_\tau(x)
&= \emx \bigg( \d^\mu \Gamma^{\rho \sigma}_\tau(x)
- m^\rho \Gamma^{\mu \sigma}_\tau(x)
- m^\sigma \Gamma^{\rho \mu}_\tau(x)  \cr
&+ \delta^\mu_\tau m^\nu \Gamma^{\rho \sigma}_\nu(x)
+ m^\rho m^\sigma \delta^\mu_\tau \bigg). }
	\eqno(3.6) $$
By means of the connection we can define the simplest kind of binary
homomorphism, the covariant derivative.

Despite their appearent simplicity, tensor fields are in a sense quite
complicated
objects. To see this we must consider their restriction to the largest
finite-dimensional
subalgebra. Morally speaking, $Vect(N)$ has a $gl(N)$ subalgebra consisting of
rigid general linear transformations. This is not strictly true, since
a vector field is an everywhere small diffeomorphism while a small linear
transformation is not small sufficiently far from the origin, but
it is true on the group level: $GL(N) \subset Diff(\Reals^N)$.
Therefore there is a close relationship between
representations of $Vect(N)$ and $gl(N)$, which
is evident in the case of tensor fields.

{}From (3.2) it follows by formal manipulations that
$$ J^\mu_\nu \equiv {{\d L^\mu(m)} \over {\d{m^\nu}}} \bigg|_{m=0}
=  x_\nu \d^\mu + T^\mu_\nu
	\eqno(3.7) $$
satisfies $gl(N)$. Note that we treat $m$ as a continuous variable, which
means that the manifold under consideration is really flat space $\Reals^N$.
The first term in (3.7) can be thought of as orbital angular
momentum (this is what it would be if $gl(N)$ were replaced by $so(N)$),
and the last term is the intrinsic ``spin''. It is now clear that
the restriction of the $Vect(N)$ module $\TT^p_q(\lambda)$ to $gl(N)$ is
$\Omega \oplus \TT^p_q(\lambda)$, where the
{\it orbital representation} $\Omega$ of $gl(N)$ is defined by
$ J^\mu_\nu = x_\nu \d^\mu.$

However, $gl(N)$ is not the largest finite-dimensional $Vect(N)$ subalgebra
(in the same moral sense as above).
We can add translations and a kind of conformal transformations to $gl(N)$
to obtain a new finite-dimensional subalgebra. The result is
related to $gl(N)$ in the same way as the ordinary conformal algebra is
related to $so(N)$, and therefore we call it the
{\it conformal linear algebra}.
The special conformal generators do not quite have the standard form,
which would require introduction of additional structure in the form of
a metric, but the kinship with the usual conformal algebra is obvious.

The conformal linear algebra $cl(N)$ is the $N(N+2)$-dimensional Lie
algebra with basis $\{P^\mu, T^\mu_\nu, K_\nu\}_{\mu,\nu = 1}^N$
and brackets
$$ \eqalign{
[P^\mu, P^\nu] &= 0 \cr
[K_\mu, K_\nu] &= 0 \cr
[J^\mu_\nu, J^\sigma_\tau] &=
\delta^\mu_\tau J^\sigma_\nu - \delta^\sigma_\nu J^\mu_\tau }
\qquad \qquad \eqalign{
[J^\mu_\nu, P^\sigma] &= - \delta^\sigma_\nu P^\mu \cr
[J^\mu_\nu, K_\tau] &= \delta^\mu_\tau K_\nu \cr
[P^\mu, K_\nu] &= \delta^\mu_\nu J^\sigma_\sigma + J^\mu_\nu. }
	\eqno(3.8) $$

It is straight-forward to check that $cl(N)$ is a Lie algebra by direct
verification of the Jacobi identities. It is even simpler to note that
it is isomorphic to $sl(N+1)$, which is the Lie algebra with basis
$\{J^A_B\}_{A,B=1}^{N+1}$, subject to the conditions
$$ [J^A_B, J^C_D] = \delta^A_D J^B_C - \delta^B_C J^A_D,
\qquad\qquad J^A_A = 0.
	\eqno(3.9) $$
The isomorphism is given by the following identifications:
$$ J^A_B \equiv
\pmatrix{J^0_0 &J^0_\nu \cr
 J^\mu_0 &J^\mu_\nu}
=
\pmatrix{- J^\sigma_\sigma &-K_\nu \cr
 P^\mu &J^\mu_\nu} ,
	\eqno(3.10) $$
where $A = (0, \mu)$, $B = (0, \nu)$ are $N+1$-dimensional indices.
Here and henceforth $N+1$-dimensional indices are denoted by capital
Latin letters from the beginning of the alphabeth.
To prove the claimed isomorphism, we must e.g. verify that
$$ \eqalign{
[J^\mu_\nu, J^\sigma_0] = [J^\mu_\nu, P^\sigma]
= -\delta^\sigma_\nu P^\mu
= \delta^\mu_0 J^\sigma_\nu - \delta^\sigma_\nu J^\mu_0, }
	\eqno(3.11) $$
because $\mu \ne 0$. The other five brackets are checked similarly.

If $L^\mu(m)$ satisfies $Vect(N)$,
the following generators satisfy $cl(N)$.
$$ P^\mu = L^\mu(0),
\qquad J^\mu_\nu = {{\d L^\mu(m)} \over {\d m^\nu}} \bigg|_{m = 0},
\qquad K_\nu = {{\d^2 L^\mu(m)} \over {\d m^\mu \d m^\nu}} \bigg|_{m = 0}.
	 \eqno(3.12) $$
This observation is the reason why $cl(N)$ is important to understand
$Vect(N)$, because it means that every $Vect(N)$ module gives rise to a
$cl(N)$ module by restriction.
In particular, from the scalar representation of $Vect(N)$ we obtain
a $cl(N)$ representation by differentiating $L^\mu(m) = \emx \d^\mu$
with respect to $m$ at $m=0$.
$$ P^\mu = \d^\mu,
\qquad J^\mu_\nu = x_\nu \d^\mu,
\qquad K_\nu = x_\nu x \ddot \d.
	\eqno(3.13) $$
In analogy with the corresponding $gl(N)$ representation,
this deserves to be called the
{\it orbital representation} of $cl(N)$ and denoted by $\Omega$.

The restriction of the tensor field $\TT^p_q(\lambda)$ to $cl(N)$ reads
$$ P^\mu = \d^\mu,
\qquad J^\mu_\nu = x_\nu \d^\mu + T^\mu_\nu,
\qquad K_\nu = x_\nu x \ddot \d + x \ddot T_\nu + x_\nu T^\sigma_\sigma.
 	\eqno(3.14) $$
whereas the connection (3.6) gives upon restriction
$$ \eqalign{
P^\mu \Gamma^{\rho \sigma}_\tau &=\d^\mu \Gamma^{\rho \sigma}_\tau \cr
J^\mu_\nu \Gamma^{\rho \sigma}_\tau &= x_\nu \d^\mu \Gamma^{\rho \sigma}_\tau
- \delta^\rho_\nu \Gamma^{\mu \sigma}_\tau
- \delta^\sigma_\nu \Gamma^{\rho \mu}_\tau
+ \delta^\mu_\tau \Gamma^{\rho \sigma}_\nu \cr
K_\nu \Gamma^{\rho \sigma}_\tau
&=  x_\nu x \ddot \d \Gamma^{\rho \sigma}_\tau
- \delta^\rho_\nu x \ddot \Gamma^\sigma_\tau
- \delta^\sigma_\nu \Gamma^\rho_\tau \ddot x
+ x_\tau \Gamma^{\rho \sigma}_\nu
- x_\nu \Gamma^{\rho \sigma}_\tau
+ \delta^\rho_\nu \delta^\sigma_\tau + \delta^\sigma_\nu \delta^\rho_\tau. }
	\eqno(3.15) $$
Note that the $gl(N)$ subalgebra is not able to distinguish the connection
from a
tensor field of type $\TT^2_1(0)$; only the last two terms in the action
of the special conformal generator achieves this.

Eq. (3.14) points at a fundamental incompleteness of tensor fields, which
motivated us to search for a new class of representations.
Since $cl(N) \cong sl(N+1)$ we know much about its representations; in
particular the irreducible finite-dimensional representations
\hbox{$\TT^p_q(\notrace)$} are $gl(N+1)$ tensors with $p$ upper and $q$
lower indices ($\lambda = \notrace$ by tracelessness).
However, restriction of
the $Vect(N)$ module $\TT^p_q(\lambda)$ does not yield any of the non-trivial
finite-dimensional $sl(N+1)$ representations, although it does yield all
$gl(N)$ modules according to (3.7). This is not surprising since tensor
fields by definition are $Vect(N)$ modules induced from $gl(N)$ tensors.
It is thus natural to ask if there are $Vect(N)$ modules whose $cl(N)$
restriction
contain finite-dimensional $sl(N+1)$ modules. The positive answer is the
main result of this paper.

\proclaim{Theorem 3.1}
The following expression satisfies $Vect(N)$.
$$ \eqalign{
L^\mu(m) &= \emx \biggl( \d^\mu + m \ddot T^\mu + (1 - m \ddot x) T^\mu_0 \cr
& \qquad + c m^\mu \bigl( m \ddot T \ddot x + m \ddot T^0
- m \ddot x \, T^0_0 - m \ddot x \, T_0 \ddot x \bigr) \biggr), } $$
where
$$ T^A_B = \pmatrix{{T^0_0} &{T^0_\nu} \cr {T^\mu_0} &{T^\mu_\nu}} $$
satisfies $gl(N+1)$.
\endproclaim

\demo{Proof}
The proof straight-forward, but since it is our main result we give
the details.
$$ \eqalign{
[L^\mu(m), &L^\nu(n)] = \emnx \biggl(
n^\mu \bigl( \d^\nu + n \ddot T^\nu + (1 - n \ddot x) T^\nu_0 \cr
&\qquad + c n^\nu \bigl( n \ddot T \ddot x + n \ddot T^0
- n \ddot x \, T^0_0 - n \ddot x \, T_0 \ddot x \bigr) - n^\mu T^\nu_0 \cr
&\qquad + c n^\nu \bigl( n \ddot T^\mu - n^\mu T^0_0 - n^\mu T_0 \ddot x
- n \ddot x \, T^\mu_0 \bigr)
+ n^\mu \, m \ddot T^\nu - (1 - n \ddot x) m^\nu T^\mu_0 \cr
&\qquad + c n^\nu \bigl( n^\mu \, m \ddot T \ddot x
- m \ddot x \, n \ddot T^\mu
+ n^\mu \, m \ddot T^0 + n \ddot x \, m \ddot x \, T^\mu_0 \bigr) \cr
&\qquad + (1 - m \ddot x) c n^\nu \bigl( n^\mu T_0 \ddot x + n^\mu T^0_0
- n \ddot T^\mu + n \ddot x \, T^\mu_0 \bigr) \cr
&\qquad + c^2 m^\mu n^\nu n \ddot x \bigl( m \ddot T \ddot x + m \ddot T^0
+ m \ddot x \, T_0 \ddot x \cr
& \qquad - m \ddot T^0 - (m \ddot T \ddot x  - m \ddot x \, T^0_0)
+ m \ddot x \, T_0 \ddot x \bigr) \biggr) - \mxn \cr
&= n^\mu \emnx \biggl( \d^\nu + (m+n) \ddot T^\nu
+ (1 - (m+n) \ddot x) T^\nu_0
+ c n^\nu \bigl( (m+n) \ddot T \ddot x \cr
&\qquad + (m+n) \ddot T^0 - (m+n) \ddot x \, T^0_0
- (m+n) \ddot x \, T_0 \ddot x \bigr) \biggr) - \mxn \cr
&= n^\mu L^\mu(m+n) - \mxn. }
	\eqno(3.16) $$
We used that $m^\nu n^\mu f(m+n) - \mxn = n^\mu (m^\nu + n^\nu) f(m+n) - \mxn$
for any function that depends on $m+n$ only. \qed
\enddemo

We claim that theorem 3.1 is the most general expression satisfying $Vect(N)$
from the following class. $L^\mu(m)$ is $\Lambda$-graded, depends on the
derivative only through $\emx\d^\mu$, and it depends otherwise only on
$m$, $x$ and the generators of $sl(N+1)$. The most general ansatz in
this class is
$$ \eqalign{
L^\mu(m) &= \emx \biggl( \d^\mu + m \ddot T^\mu + \alpha(m \ddot x) T^\mu_0
+ m^\mu\bigl( \beta(m \ddot x) m \ddot T \ddot x \cr
&\qquad + \gamma(m \ddot x) m \ddot T^0 + \epsilon(m \ddot x) T^0_0
+ \phi(m \ddot x) T_0 \ddot x \bigr) \biggr), }
	\eqno(3.17) $$
where $\alpha, \ldots, \phi$ are functions of $m \ddot x$ and
$T^A_B \in gl(N+1)$. Note that no term proportional to $T^\sigma_\sigma$ is
included because it equals $-T^0_0$ in $sl(N+1)$.
When this ansatz is inserted into the brackets, a slightly more general
expression than that in the theorem turns out to be consistent. However,
if $T^A_B$ satisfies $gl(N+1)$, so does
$$ T^{\prime A}_B =
\pmatrix{T^0_0 &T^0_\nu / \alpha \cr \alpha T^\mu_0 &T^\mu_\nu }.
	\eqno(3.18) $$
Once this freedom is eliminated the expression above results.

It should be emphasized that although our motivation for the inadequacy of
tensor fields depends on differentiation with respect to $m$, which is a
quite formal manipulation, the result in theorem 3.1 does not.
To better understand the nature this result it is useful to introduce
an $N+1$-dimensional formalism, with momenta $m^A$ and coordinates $x_B$,
by the following definitions
$$ m^A \equiv (m^0, m^\mu) = (-m \ddot x, m^\mu), \qquad
x_B \equiv (x_0, x_\nu) = (1, x_\nu).
	\eqno(3.19) $$
It is clear that $m^A x_A \equiv 0$. Moreover, we indicate contraction of
$(N+1)$-dimensional indices by double dots: $m\:x \equiv m^A x_A$.
A single dot indicates $N$-dimensional contraction, as before:
$m \ddot x \equiv m^\mu x_\mu$. Thus theorem 3.1 acquires the form

\proclaim{Theorem 3.2}
The following expression satisfies $Vect(N)$.
$$ L^\mu(m) =
\emx \bigl( \d^\mu + T^\mu_0 + m\:T^\mu + c m^\mu m\:T\:x\bigr), $$
where $T^A_B \in gl(N+1)$ and
$$ \eqalign{
m\:x &\equiv 0, \qquad
[\d^\mu, n^A] = - \delta^A_0 n^\mu, \qquad
[\d^\mu, x_B] = \delta^\mu_B, \qquad
x_0 = 1. } $$
\endproclaim

\demo{Proof}
$$ \eqalign{
[L^\mu(m), &L^\nu(n)]
= \bigl[ \emx \bigl( \d^\mu + T^\mu_0 + m\:T^\mu
 + c m^\mu m\:T\:x\bigr), \cr
&\qquad \enx \bigl(\d^\nu + T^\nu_0 + n\:T^\nu + c n^\nu n\:T\:x\bigr)
 \bigr] \cr
&= \emnx \biggl( n^\mu(\d^\nu + T^\nu_0 + n\:T^\nu + c n^\nu n\:T\:x)
 - n^\mu T^\nu_0 \cr
& \qquad + c n^\nu( - n^\mu T_0\:x + n\:T^\mu) + n^\mu T^\nu_0
+ c n^\nu(n^\mu T_0\:x - n\:T^\mu x_0) \cr
& \qquad + n^\mu m\:T^\nu + c n^\nu(n^\mu m\:T\:x - m\:x \, n\:T^\mu) \cr
& \qquad+ c^2 m^\mu n^\nu n\:x \, m\:T\:x \biggr) - \mxn \cr
&= n^\mu \emnx \bigl( \d^\nu + T^\nu_0 + (m+n)\:T^\nu
+ c n^\nu (m+n)\:T\:x\bigr) - \mxn \cr
&= n^\mu L^\nu(m+n) - \mxn. }
	\eqno(3.20) $$
We used that $m\:x = n\:x = 0$. \qed
\enddemo

It is now clear that we can build new $Vect(N)$ modules by considering the
action given by theorem 3.1 on elements of the form
$$ \phi(x) = \upsilon \otimes f(x).
	\eqno(3.21) $$
where $\upsilon$ is a $\TT^p_q(\lambda)$ $gl(N+1)$ tensor and $f(x)$ is a
scalar function. We denote this $Vect(N)$ module by $\CC^p_q(\lambda, c)$.
As examples we write down the action of $Vect(N)$ on
$\CC^1_0(0, c)$, $\CC^0_1(0, c)$ and $\CC^0_0(\lambda, c)$.
$$ \eqalign{
L^\mu(m) \phi^A &= \emx \bigl( \d^\mu \phi^A
- \delta^A_0 \phi^\mu - m^A \phi^\mu - c m^\mu m^A \, x\:\phi \bigr) \cr
L^\mu(m) \phi_B &= \emx \bigl( \d^\mu \phi_B
+ \delta^\mu_B \phi_0 + \delta^\mu_B m\:\phi
+ c m^\mu x_B \, m\:\phi \bigr) \cr
L^\mu(m) \phi &= \emx \bigl( \d^\mu \phi
+ \lambda m^\mu \phi + c m^\mu (\lambda \, m\:x)\phi \bigr) \cr
&= \emx \bigl( \d^\mu \phi + \lambda m^\mu \phi \bigr). }
	\eqno(3.22) $$
It is clear that $\CC^0_0(\lambda, c) \cong \TT^0_0(\lambda)$.

By differentiation of the expression in theorem 3.1 with respect to $m$, it
is found that the expressions in theorems 3.1-2 correspond to the
following $cl(N)$ generators.
$$ \eqalign{
P^\mu &= \d^\mu + T^\mu_0 \cr
J^\mu_\nu &= x_\nu \d^\mu + T^\mu_\nu \cr
K_\nu &= x_\nu x \ddot \d + x_\nu T^\sigma_\sigma - x_\nu T_0 \ddot x
+ T_\nu \ddot x + c(N+1)(T_\nu\:x - T_0\:x \, x_\nu). }
	\eqno(3.23) $$
The expression for $K_\nu$ reads in $N$-dimensional notation
$$ \eqalign{
K_\nu &= x_\nu x \ddot \d + c(N+1) T^0_\nu
+ x_\nu (T^\sigma_\sigma - c(N+1) T^0_0) \cr
&\qquad -(1+c(N+1)) x_\nu T_0 \ddot x
+(1+c(N+1)) T_\nu \ddot x, } \eqno(3.24) $$
and particularly when $c=-1/(N+1)$ and $T^A_A = 0$,
$$ K_\nu = x_\nu x \ddot \d - T^0_\nu. \eqno(3.25)  $$

{}From the last expression, it is clear that the restriction of
\hbox{$\CC^p_q(\notrace, -1/(N+1))$} to $sl(N+1)$
is $\Omega \oplus \TT^p_q(\notrace)$.
This means that these modules give rise to all finite-dimensional
representations of the conformal algebra $sl(N+1)$ upon restriction, which
would motivate to name them {\it conformal fields}.

Let us finally discuss on the physical meaning of some of the parameters
characterizing conformal fields. To this end we make some simple
observations.
$P^\mu$ is the generator of rigid translations, i.e. the
momentum operator, and we can therefore identify its eigenvalue in a
$P^\mu$ eigenstate with the momentum of this state. Since
$P^\mu = \d^\mu + T^\mu_0$, $T^\mu_0$ takes the role of a characteristic
momentum, and it is tempting to identify an eigenvalue of
$T^N_0$ as a mass, $N$ being the time direction.
The eigenstates of the dilatation operator
$J^\mu_\mu = x \ddot \d + T^\mu_\mu $
are scale invariant, wherefore the eigenvalues of $T^\mu_\mu $
should give critical exponents, which
could be relevant to critical systems in $N$ dimensions.

\endchapter

\header{4. Fock modules}

In this section we discuss the construction of Fock modules for
$Vect(N)$, and
substantiate the claim in section 2 that infinite central extensions arise.
Any Fock module, or more generally any lowest-weight module,
is characterized by a $\Ints$-gradation and a lowest-weight state of
minimal degree.
We focus our attention to modules whose gradation is by one component of the
momentum. If this is the time component, the gradation is by energy and
the lowest weight can be thought of as a mass. Other $\Ints$-gradation are
clearly possible, e.g. according to the value of the dilatation operator.

In a Fourier transformed basis, a tensor field is a $Vect(N)$ module
with basis $\{\phi(n)\}_{n \in \Lambda}$ and action
$$ L^\mu(m) \phi(n) = (n^\mu + m \ddot T^\mu) \phi(m+n)
	\eqno(4.1) $$
A slightly more general representation is found by considering
$\psi(n) = \phi(n+h)$,
$h$ a constant vector. The action follows from (4.1) by replacing
$n^\mu$ by $n^\mu + h^\mu$. Of course, $\psi$ and $\phi$ are related
by a change of
basis if $h \in \Lambda$, so $h$ is only defined modulo $\Lambda$. It is
tempting to interpret $h$ as a ``mass'', or rather as a characteristic
momentum which might point in the time direction.
Note that this ``mass'' is related to conformal weights of primary
Virasoro fields.

Let $a(m)$ and $\bar a(n)$ be bosonic oscillators,
satisfying the canonical commutation relations
$$ [a(m), \bar a(n)] = \delta(m+n), \qquad
[a(m), a(n)] = [\bar a(m), \bar a(n)] = 0.
	\eqno(4.2) $$
Then the following expression satisfies $Vect(N)$.
$$ L^\mu(m) = - \sum_{s\in\Lambda} \bar a(m-s)
\, (s^\mu + h^\mu + m \ddot T^\mu) \, a(s),
	\eqno(4.3) $$
and the action of (4.3) on $a(n)$ is the shifted variant of (4.1).
This representation extends naturally to arbitrary polynomials in $a(n)$,
i.e. symmetrized tensor powers of (4.1).
Because (4.3) commutes with the bosonic number operator
$$\sum_{s\in\Lambda} \bar a(m-s) \, a(s),
	\eqno(4.4) $$
each monomial in $a$ is closed under the action of $Vect(N)$.

To make this into a Fock module, we introduce a division of the
lattice $\Lambda$,
$$ \Lambda = \Lambda_- \cup \{0\} \cup \Lambda_+,
	\eqno(4.5) $$
and write $m > 0$ ($m < 0$) if $m \in \Lambda_+$ ($m \in \Lambda_-$).
The decomposition must be such that $m, n > 0$ implies that $m+n > 0$ and
$-m < 0$.
A division of this kind can e.g. be defined by introducing a constant vector
$k_\mu$: $m > 0$ iff $k \ddot m > 0$. If there are non-zero points
satisfying $k \ddot m = 0$ some extra effort has to be taken to divide
these points
equally between the two halves. We can now define a {\it vacuum state}
$\vacuum$ by $\bar a(0) \vacuum = 0$ and
$a(m) \vacuum = \bar a(m) \vacuum = 0$ for all $m < 0$. This means that
$ L^\mu(m) \vacuum $ also vanishes for all $m < 0$, where $L^\mu(m)$ given
by (4.3),
because either $a(s) < 0$ or $ \bar a(m-s) < 0$ and the two commute.
However, the action of $L^\mu(0)$ diverges.
$$ L^\mu(0)\, \vacuum
= - \sum_{s\in\Lambda} \bar a(-s) \, (s^\mu + h^\mu) \, a(s) \, \vacuum
= \sum_{s > 0} (s^\mu + h^\mu) \, \vacuum.
	\eqno(4.6)$$

When $N = 1$, the standard approach to avoid this infinity is normal ordering,
but that idea does not work for $N > 1$. To see what goes wrong, consider
the case
$h^\mu = T^\mu_\nu = 0$. The only normal ordered generators differing from
(4.6) are
$$ L^\mu(0) = - \sum_{s<0} \bar a(-s) \, s^\mu \, a(s)
- \sum_{s>0} a(s) \, s^\mu \, \bar a(-s),
	\eqno(4.7) $$
When computing $[L^\mu(m), L^\nu(n)]$ we pick up a central term
proportional to
$\delta(m+n)$, and the proportionality constant is
$$ \sum_{0 < s < m} s^\mu (m^\nu - s^\nu).
	\eqno(4.8)$$
In one dimension this sum can be readily performed, yielding $(m^3 - m)/6$
($c = -2$), but when
$N \ge 2$ the set of points between $0$ and $m$ is infinite. More precisely,
the sum is proportional to $\infty^{N-1}$, where $\infty$ is the number of
integers. Of
course, this is a signal that the normal ordering prescription breaks down,
which is
in accordance with the result of section 2 that there is no central extension.

The same divergence has been noted by Figueirido and Ramos\ref{9}, who draw
the bold
conclusion that the Jacobi identities have to be abandoned. We propose a
less drastic
way out. Note that the same steps can be repeated with fermionic oscillators,
satisfying canonical anti-commutation relations
$$ \{b(m), \bar b(n)\} = \delta(m+n), \qquad
\{b(m), b(n)\} = \{\bar b(m), \bar b(n)\} = 0,
	\eqno(4.9) $$
and the $Vect(N)$ generators given by (4.3) with all $a$'s replaced by $b$'s.
We find that
$$ L^\mu(0)\vacuum = -\sum_{s > 0} (s^\mu + h^\mu) \, \vacuum.
	\eqno(4.10) $$
If we now consider a theory with $N_B$ bosonic and $N_F$ fermionic species,
with ``masses'' $h_i$ and $h_j$, respectively, the total vacuum eigenvalue
becomes
$$ L^\mu(0)\vacuum = \sum_{s > 0} \bigg( (N_B - N_F) s^\mu
+ (\sum_{i=1}^{N_B} h_i^\mu - \sum_{j=1}^{N_F} h_j^\mu) \bigg) \vacuum.
 	\eqno(4.11)$$
This expression vanishes, in spite of the divergent sum over $s$, provided
that
$N_F = N_B$ and $\sum h_i^\mu = \sum h_j^\mu$.

Eq. (4.11) is an expression of type $\infty \times 0$, which in general could
be anything, but we argue that it must vanish for the following reason.
If the sum
over $s$ is cut off at large momenta, we obtain generators which obey some
approximation to $Vect(N)$. In the limit that the cutoff approaches infinity,
this approximation should be increasingly good. However, (4.11) is
identically zero for every finite cutoff, and thus the limit is also zero.
Note that the result was formulated for tensor fields, but
it also holds for conformal fields, because $L^\mu(0)$ is of the
form (4.6) with $h^\mu = T^\mu_0$.

The impossibility of normal ordering is not a peculiarity of the plane
wave basis. We can write down an expression
for the generators using a position space basis,
$$ L^\mu(m) = - \int d^N\!x \, \emx \, \bar a(x) \,
(\d^\mu + h^\mu + m \ddot T^\mu) \, a(x),
	\eqno(4.12) $$
where
$$ [a(x), \bar a(y)] = \delta(x-y),
	\eqno(4.13) $$
The oscillators can be expanded as
$$ a(x) = \sum_{\ii} a_{\ii} \, \varphi_\ii(x),
\qquad \bar a(x) = \sum_{\ii} \bar a_{\ii} \, \varphi_\ii(x),
	\eqno(4.14) $$
where $\{\varphi_\ii\}_{\ii\in I}$ is a complete orthogonal function
basis and
$I$ is an index set. We can now define a total order based upon the first
component
of $\ii$ and a corresponding Fock module. Divergences arise because the
other components of $\ii$ can take infinitely many values, except in one
dimension where a single index suffices to label a complete set of functions.
In particular we can use the sperical basis
$x = (r, \Omega)$, $\ii = (n, \ll)$ and $ \varphi_\ii(x) = r^n Y_\ll(\Omega)$,
where $Y_\ll$ is the $N$-dimensional sperical harmonics. In this way we
obtain Fock modules graded according to the dilatation eigenvalue.

To summarize, the vacuum eigenvalue does not diverge
provided that the number of bosonic and fermionic degrees of freedom
are the same, as well as the total bosonic and fermionic ``masses''. This
condition is slightly reminiscent of supersymmetry, but it is not equivalent.
The $Vect(N)$ generators do not have any fermionic partners, and
the bosonic and fermionic number operators still commute with $L^\mu(m)$,
wherefore Fock modules decompose into sectors with a fixed number of
particles. The situation is thus somewhat paradoxical;
bosons and fermions do not transform into each other, but they interact in
a subtle way through the vacuum to remove infinities.

The Fock space construction is completely general and can be applied to other
multi-graded Lie algebras, e.g. $Map(N, \oj)$, the algebra of maps from
$N$-dimensional space to a finite-dimensional Lie algebra $\oj$. If $M^a$
are matrices in a finite-dimensional representation of $\oj$,
$$ T^a(m)  = \sum_{s\in\Lambda} \bar a(m-s) \, M^a \, a(s),
	\eqno(4.15) $$
satisfies $Map(N, \oj)$ (representation indices are suppressed). If this
expression
is normal ordered, one picks up a central extension proportional to the
number of
points between $0$ and $m$, i.e. infinity. Again, this infinity could be
cancelled against a fermionic contribution.

Eq. (4.3) defines a quadratic embedding of $Vect(N)$ in an infinite Heisenberg
algebra. There are two other important quadratic embeddings of $Vect(1)$: in
Kac-Moody algebras (Sugawara construction) and in the algebra of bosonic
string
oscillators\ref{14}. However, it is easy to see that neither of these
constructions
have any higher-dimensional counterpart, even on the classical level, because
the index structure would be wrong. The
Sugawara construction would be something like
$$ L^\mu(m) = \sum_{s\in\Lambda} T^a(m-s) \, T^a(s),
	\eqno(4.16) $$
which does not make sense because there is vector to the left and a scalar
to the right. String oscillators should satisfy
$$ [a(m), a(n)] = m^\mu \, \delta(m+n),
	\eqno(4.17) $$
but since $m^\mu$ has a vector index $a(m)$ would have to be ``half-vector'',
which we do not know how to treat.

\endchapter

\header{5. Discussion}

The results in section 3 can be used to construct representations of
the algebra of Poisson brackets in a $N$-dimensional phase space ($N$ even).
It is given by the brackets
$$ [f, g] = \omega_{\mu\nu} \, \d^\mu f \, \d^\nu g,
	\eqno(5.1) $$
where $\omega_{\mu\nu}$ is the constant, anti-symmetric, non-degenerate
symplectic form. By expanding the functions in the plane-wave basis
$\{E(m) \equiv \emx\}_{m \in \Lambda}$, we obtain
$$ [E(m), E(n)] = \omega_{\mu\nu} m^\mu n^\nu E(m+n).
	\eqno(5.2) $$
The adjoint representation of (5.2) is given by
$$ E(m) = \emx \omega_{\mu\nu} m^\mu \d^\nu.
	\eqno(5.3) $$
We now note that the defining $Vect(N)$ representation (2.4) is given by
$ L^\mu(n) = \emx \d^\mu$, and thus
$$ E(m) = \omega_{\mu\nu} m^\mu L^\nu(m)
	\eqno(5.4) $$
satisfies (5.2), provided that $L^\mu(n)$ is in the defining representation.
However, it is easy to check that a sufficient condition for the expression
(5.4) to satisfy (5.2) is that $L^\mu(n)$ obeys (2.2) ($L^\mu(m)$ commutes
with $\omega_{\sigma\tau}$), and hence we can insert any $Vect(N)$
representation in
(5.4) to obtain a new representation of the Poisson algebra. Together with the
results of section 3 this gives many new representations.

The Poisson algebra admits a Lie algebra deformation, the Moyal algebra, which
takes the form
$$ [E(m), E(n)] = (\e^{i \hbar \omega_{\mu\nu} m^\mu n^\nu} -
\e^{-i \hbar \omega_{\mu\nu} m^\mu n^\nu})  E(m+n).
	\eqno(5.5) $$
After a trivial rescaling of the generators, the $\hbar \to 0$ limit of
(5.5) is
clearly (5.2). Replacing the Poisson algebra by the Moyal algebra is one route
to quantization, advocated by Bayen et al.\ref{17} (see also Ref. 18).
It is now natural to
ask if (5.4) also can be deformed, i.e. if one can make the substitution
$\d^\mu \to \enmx L^\mu(m)$ in the adjoint representation
$$ E(m) = (\e^{i \hbar \omega_{\mu\nu} m^\mu \d^\nu} -
\e^{-i \hbar \omega_{\mu\nu} m^\mu \d^\nu}).
	\eqno(5.6) $$
The answer is negative; it seems impossible to generalize (5.4) to the
Moyal algebra.

We hope that the new representations of $Vect(N)$ discovered in this
paper could eventually have some applications to physics. A rather
obvious field is quantum gravity, which almost by definition is
intimately related to action of the diffeomorphism group.  It is safe
to say that quantized gravity theories based on tensor fields have not
been very successful; conformal fields may fare better.  Another idea
could be to look for a classification of $N$-dimensional phase
transitions, similar to conformal field theory in two dimensions.  Of
course, the diffeomorphism group is much bigger than the conformal
group, even in two dimensions, so this would require much more than a
direct generalization of conformal field theory to $N$ dimensions.  On
the other hand, unless we wish to consider artefacts of the choice of
coordinate system, it is hard to see how arbitrary diffeomorphisms can
fail to be a symmetry of any sensible theory. However, such a theory
will presumably include gravity.

\vskip 20pt
\header{Acknowledgements}

I am grateful to Bengt Nagel for helpful comments.

\newpage

\noindent\header{References}

\noindent 1. A. A. Kirillov, Russian Math. Surveys 31:4 (1976) 57.

\noindent 2. A. M. Vershik, I. M. Gelfand and M. I. Graev,
Russian Math. Surveys 30:6 (1975) 1.

\noindent 3. A. A. Rudakov, Math. USSR Izvestija 8 (1974) 836.

\noindent 4. J. N. Bernstein and D. A. Leites, Sel. Math. Sov. 1
(1981) 143.

\noindent 5. D. B. Fuks, Cohomology of infinite-dimensional Lie algebras
(New York and London: Plenum Press 1987).

\noindent 6. R. Ree, Trans. Amer. Math. Soc. 83 (1956) 510.

\noindent 7. E. Ramos and R. E. Shrock, Int. J. Mod. Phys. A 4 (1989) 4295.

\noindent 8. E. Ramos, C. H. Sah and R. E. Shrock, J. Math. Phys. 31
(1989) 1805.

\noindent 9. F. Figueirido and E. Ramos, Int. J. Mod. Phys. A 5 (1991) 771.

\noindent 10. T. A. Larsson, Phys. Lett. B 231 (1989) 94.

\noindent 11. T. A. Larsson, to appear in J. Phys. A (1992).

\noindent 12. E. Ragoucy and P. Sorba, preprint CERN-TH-5737 (1990).

\noindent 13. A. Pressley and G Segal, Loop groups,
(Oxford: Claredon Press 1986).

\noindent 14. P. Goddard and D. Olive, Int. J. Mod. Phys. A 1 (1986) 303.

\noindent 15. B. L. Feigin and D. B. Fuks, Funct. Anal. and Appl. 16
(1982) 144.

\noindent 16. A. A. Belavin, A. M. Polykov and A. B. Zamolodchikov,
Nucl. Phys. B 241 (1984) 333.

\noindent 17. F. Bayen, M. Flato, C. Fronsdal, A. Lichnerowicz and
D. Sternheimer, Annals of Physics 111 (1978) 61, 111.

\noindent 18. D. B. Fairlie, P. Fletcher and C. K. Zachos,
Phys. Lett. B 218 (1989) 203.

\vfill
\end